\documentclass[11pt,onecolumn,letterpaper]{IEEEtran}

\usepackage{amsmath,amsfonts,amssymb}
\usepackage{graphicx}
\title{A Simple Converse of Burnashev's Reliability Function}
\author{\authorblockN{Peter Berlin$^\dagger$,  Bar\i\c{s}
Nakibo\u{g}lu{$^\ddagger$}, Bixio Rimoldi$^\dagger$, Emre Telatar$^\dagger$}\\
\authorblockA{\vspace{0.3\baselineskip}$^\dagger$School of Computer and
Communication Sciences\\ Ecole Polytechnique F\'ed\'erale de Lausanne (EPFL)\\
CH-1015 Lausanne, Switzerland\\
\vspace{0.3\baselineskip}{$^\ddagger$} Department of Electrical Engineering and Computer Science\\ 
 Massachusetts Institute ofTechnology (MIT)\\MA 02139 Cambridge, USA\\ \vspace{0.3\baselineskip}
\texttt{peter.berlin@epfl.ch, nakib@mit.edu,\\bixio.rimoldi@epfl.ch,
emre.telatar@epfl.ch}} }

\newtheorem{theorem}{Theorem} 
\newtheorem{proposition}{Proposition}
 
\newtheorem{lemma}{Lemma}
\newcommand{\DEF}{{\triangleq}}

\newcommand{\cG}{{\cal G}}
\newcommand{\cGt}{{\cal G}{(Y^{\tau})}}

\newcommand{\cS}{{\cal S}} 
\newcommand{\cT}{{\cal T}} 

\newcommand{\cW}{{\cal W}}
\newcommand{\cX}{{\cal X}}
\newcommand{\cY}{{\cal Y}}
\newcommand{\cZ}{{\cal Z}}

\newcommand{\Pe}[0] {{\it P_{e}}}     %Probability of error
\newcommand{\PRM}[1]{{\cal Q}_{#1}}
\DeclareMathOperator{\PrA}{P_{A}}
\DeclareMathOperator{\PrN}{P_{N}} 
\newcommand{\PX}[1]{\Pr\left\{{#1}\right\} }  %Probability 
\newcommand{\EX}[1]{E\left[{#1}\right]} %Expectation

\newcommand{\h}[1]{ {\mathcal H}(W|#1)}
\newcommand{\HX}[1]{\h{Y^{{#1}}}} 

\newcommand{\PCX}[2]{\PX{ \left.\! {#1} \right| {#2} }} %Conditional Probability 
\newcommand{\ECX}[2]{\EX{ \left.\! {#1} \right| {#2} }} %Conditional Expectation 
\newcommand{\KLD}[2]{D\left( \left.\! {#1} \right\|{#2} \right)}
\newcommand{\CX}{C} 
\newcommand{\DX}{C_1} 

\DeclareMathOperator{\binh}{h}

\newcommand{\latin}[1]{\emph{#1}}

\begin{document} 
\maketitle

\begin{abstract} 
  \noindent In a remarkable paper published in 1976, Burnashev
  determined the reliability function of variable-length block codes over 
  discrete memoryless channels with feedback. Subsequently, an alternative 
  {\em achievability} proof was obtained by Yamamoto and Itoh via a particularly simple and instructive scheme. 
Their idea is to alternate between a communication and a confirmation phase until the receiver detects the codeword used by the sender to acknowledge that the message is correct.  We provide a {\em converse} that parallels the Yamamoto-Itoh achievability 
construction. Besides being simpler than the original, the proposed converse suggests that
a communication and a confirmation phase are implicit in any scheme for which the
probability of error decreases with the largest possible exponent. The proposed converse also 
makes it intuitively clear why  the terms   that appear in Burnashev's exponent are necessary.
\end{abstract}


\section{Introduction}
It is well known~(see e.g. \cite{Sha56} and \cite{Csi73}),  that the capacity of a discrete memoryless channel (DMC) is not increased by feedback.\footnote{According to common practice, we say that feedback is available if the encoder may select the current channel input as a function not only of the message but also of all past channel outputs.}   Nevertheless, feedback can help in at least two ways: for a fixed target error probability, feedback  can be used to reduce the  sender/receiver complexity and/or to reduce the expected decoding delay.  An example is the binary erasure channel, where feedback makes it possible to implement a communication strategy that is extremely simple and also minimizes the delay. The strategy is simply to send each information bit repeatedly until it is received unerased. This strategy is capacity achieving, results in zero probability of error, and reproduces each information bit with the smallest delay among all possible strategies. 

The reliability function---also called the error exponent---is a natural way to quantify the benefit of feedback.  For block codes on channels without feedback the reliability function is defined as
\begin{equation}
  E(R)=\limsup_{T\to\infty} -\frac1T \ln P_e(\lceil e^{RT}\rceil, T),
\end{equation}
where $P_e(M,T)$ is the smallest possible error probability of length $T$ block codes with $M$ codewords.
%\footnote{Strictly speaking, in the absence of feedback it is not known if the limit exists for all rates.  However, the  $\limsup$ is upper-bounded by the sphere packing bound and the $\liminf$ is lower-bounded by  the random coding bound.}

The decoding time $T$ in a communication system with feedback may depend on the channel output sequence.\footnote{ If the decoding time is not fixed, in the absence of feedback the sender may not know when the receiver has decoded.  This problem does not exist if there is feedback. }
If  it does, the decoding time $T$ becomes a random variable and the notions of rate and reliability function need to be redefined. Following Burnashev~\cite{Bur76}, in this case we define the rate as
\begin{equation}
   R \DEF  \frac{\ln M}{\EX{T}},
\end{equation}
where $M$ is the size of the message set. Similarly we  define the reliability function as
\begin{equation} 
  E_f(R) \DEF \lim_{t\to\infty} -\frac1{t}\ln
  P_{e,f}(\lceil e^{Rt}\rceil, t) ,
\end{equation}
where $P_{e,f}(M,t)$ is the smallest error probability of a variable-length block code with feedback that transmits one of $M$ equiprobable messages by means of $t$ or fewer channel uses on average. As we remark below, the limit exists for all rates from zero to capacity. 

\begin{figure}[htbp]
\centering
\begin{minipage}{0.8\textwidth} \centering \setlength{\unitlength}{0.6bp}
	  \begin{picture}(220,270)
		\put(0,0){\includegraphics[scale=0.6]{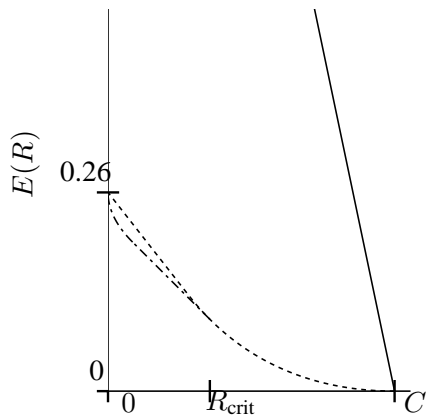}}
% Rate axis labels
\put(20,0){\makebox(0,0){$0$}} \put(84,0){\makebox(0,0){$R_\text{crit}$}}
\put(200,0){\makebox(0,0){$C$}}
% Exponent Axis labels
\put(-45,150){\makebox(0,0){\rotatebox{90}{$E(R)$}}}
\put(0,20){\makebox(0,0){0}} \put(-7,145){\makebox(0,0){0.26}}
%\put(-7,270){\makebox(0,0){0.51}}
\end{picture} \caption{Reliability functions for a binary symmetric channel with
crossover probability 0.1. Shown is Burnashev's reliability function for channels with
feedback (solid line) and upper and lower bounds to the reliability function for channels
without feedback.  The upper bound (dotted line) is given by the \emph{straight line
bound} at low rates and by the \emph{sphere packing bound} at higher rates. The lower bound
(dot-dashed line) is given by the \emph{expurgated bound}. The upper and lower bounds coincide above the critical rate, $R_\text{crit}$.} \label{fig:exponents}
\end{minipage} 
\end{figure}

Burnashev showed that for a DMC of capacity $C$, the reliability function $E_f(R)$ equals \begin{equation}
\label{eq:bexp}
 E_B(R) = C_1(1-R/C),\quad 0\leq R\leq C,
\end{equation}
where $C_1$ is
determined by the two ``most distinguishable'' channel input symbols as 
\[ C_1=\max_{x,x'} D(p(\cdot|x)\|p(\cdot|x')),  \]
where $p(\cdot|x)$ is the probability distribution of the channel output when the input is $x$, and $D(\cdot\|\cdot)$ denotes the 
Kullback-Liebler divergence between two probability distributions.  It is remarkable
that (\ref{eq:bexp}) determines the reliability function exactly for all rates. In contrast, the reliability function without feedback is known exactly only for rates above a critical rate. Below the critical rate only upper and lower bounds to the reliability function without feedback are known. For a binary symmetric channel the situation is depicted in Fig.~\ref{fig:exponents}.  

Burnashev showed that $E_f =  E_B$ by showing that for every communication scheme
\begin{equation}
  \label{eq:whatweprove} E[T]\geq \biggl(\frac{\ln M}{C} - \frac{\ln   P_e}{C_1}\biggr) \bigl(1-o(1)\bigr) 
\end{equation}
where $o(1)$ represents positive terms that tend to zero as $1/P_e$ tends to infinity,
and that there exists schemes with
\begin{equation}
 \label{eq:ach} E[T]\leq \biggl(\frac{\ln M}{C} - \frac{\ln   P_e}{C_1}\biggr) \bigl(1+o(1)\bigr),
\end{equation}
where  $o(1)$ now  represents positive terms that tend to zero as both $M$ and $1/P_e$ tend to infinity.

For a plausibility argument that justifies  (\ref{eq:ach}) it suffices to summarize the achievability construction by Yamamoto and Itoh~\cite{YamIto79}. Their scheme relies on two distinct transmission phases  that  we
shall call the communication and the confirmation phase, respectively.   
In the communication phase the message is encoded
using a fixed-length block code and the  codeword is transmitted over the forward channel. The
decoder makes a tentative decision based on the corresponding channel output. The encoder knows the channel output and can run the algorithm used by the receiver to determine the tentative decision. If the tentative decision is correct, in the confirmation phase the encoder
sends ACK. Otherwise it sends   NACK. ACKs and NACKs 
are sent via a fixed-length repetition code. (The code consists of two codewords). During the 
confirmation phase the
decoder performs a binary hypothesis test to decide if ACK or NACK was
transmitted. If ACK is decoded, the tentative decision becomes final and the transmission
of the current message ends, leaving the system free to restart with a new message.
If NACK is decoded, the tentative decision is discarded
and the two phase scheme restarts with the same message.  

The overhead caused by retransmissions is negligible if the probability of decoding NACK is small. This is the case
if both the error probability of the communication phase as well as that of the confirmation phase are 
small. Assuming that this is the case, the number of channel uses for the communication phase (including repetitions) is 
slightly above  $(\ln M)/C$. The probability of error is the probability that NACK is sent and ACK is decoded. In the asymptotic regime of interest this probability is dominated by  the probability that ACK is decoded given that NACK is sent. In a straightforward application of Stein's  lemma~\cite{Che56} one immediately sees that we can make this probability to be slightly less   than $P_e$ (thus achieve error probability $P_e$) by means of a confirmation code of length slightly above $(-\ln P_e)/ C_1$.  Summing up, we see that we can make the error probability arbitrarily close to $P_e$ by means of slightly more than $(\ln M)/C-(\ln P_e)/ C_1$  channel uses on average. This confirms (\ref{eq:ach}).

To obtain the converse   (\ref{eq:whatweprove}), Burnashev investigated the entropy of the  \latin{a posteriori} probability distribution over the message set. He showed that the average decrease of this entropy due to an additional channel output observation, as well as the average decrease of the logarithm of this entropy, are bounded.
%
%\footnote{Burnashev initially proved this bound only for small values of entropy and published \cite{Bur76} with a proof of the converse using that. However, he gave the sketch of a less complicated proof which relies on the conjecture that this bound can be established for all values of entropy. Later in \cite{Bur77} he showed that the bound holds for all values of the entropy of the posterior.} 
% Burnashev's proof is an excellent example of the power of martingales. His proof is not elementary though. 
He uses these bounds to form two submartingales, one based on the entropy of the \latin{a posteriori} distribution and the other based on the logarithm of this entropy. He then constructs a single submartingale by patching these two together.  Then Doob's optional stopping theorem is applied to this submartingale and the desired bound on the expected decoding time, which is a stopping time, is obtained. Burnashev's proof is an excellent example of the power of martingales, however both the sophistication of the martingale  construction and the use of the logarithm of entropy leaves the reader with little insight about some of the terms in the converse bound. While it is easy to see that $(\ln M)/C$ channel uses are needed in average, it was not as clear why one needs an additional  $(-\ln P_e)/ C_1$ channel uses. The connection of the latter term to binary hypothesis testing suggested the existence of an operational justification. The work presented in this paper started as an attempt to find this operational justification.

Our converse somewhat parallels the Yamamoto--Itoh achievability scheme. This suggests that that a communication and confirmation phase may be implicit components of any scheme for which the probability of error decreases with the largest possible exponent. Our approach has been generalized by Como, Y\"uksel and Tatikonda in  \cite{ComYuk07} to prove a similar converse for variable-length block codes on Finite State Markov Channels.

\section{Channel Model and Variable-Length Codes as Trees}
We consider a discrete memoryless channel, with finite input alphabet $\cX$, finite  output alphabet $\cY$, and transition probabilities $p(y|x)$. We will denote the channel input and output symbols at time $n$ by $X_n$ and $Y_n$, and denote the corresponding vectors $(X_1,X_2,\ldots,X_n)$ and $(Y_1,Y_2,\ldots,Y_n)$ by $X^n$ and $Y^n$, respectively. A perfect causal feedback link is available, i.e., at time $n$ the encoder knows $y^{n-1}$. (Following common practice, random variables are represented by capital letters and their realizations are denoted by the corresponding lowercase letters.)
 
We will assume, without loss of generality, that the channel has no ``useless
outputs symbols'', i.e., no symbols $y$ for which $p(y|x) = 0$ for every $x$.
Note that for channels for which $C_1$ is infinite, the lower bound to the
expected decoding time is a restatement of the fact that feedback does not
increase capacity.  We will therefore restrict our attention to channels for
which $C_1<\infty$.  For such channels, $p(y|x)>0$ for every $x$ and $y$; if
not, there exists an $x$ and $y$ for which $p(y|x)=0$.  Since $y$ is reachable
from some input, there also exists an $x'$ for which $p(y|x')>0$.  But then
$D(p(\cdot|x')\|p(\cdot|x))=\infty$ contradicting the finiteness of $C_1$.  The
fact that both $\cal X$ and $\cal Y$ are finite sets lets us further conclude
that for the channels of interest to this paper, there is a $\lambda>0$ for
which $p(y|x)\geq\lambda$ for every $x$ and $y$.

A variable-length block code is defined by two maps: the encoder and the decoder. The encoder\footnote{ For clarity of exposition we will only treat  deterministic coding strategies here. Randomized strategies  may be included without significant modification to the core of the proof.} functions $f_n(\cdot,\cdot):  \cW \times {\cY}^{n-1} \rightarrow \cX$, where $\cW = \{1,\ldots,M\}$ is the set of all possible messages, determine the channel input $X_n=f_n(W,Y^{n-1})$ based on the message $W $ and on past channel outputs $Y^{n-1}$. The decoder function $\hat{W}(\cdot): \cZ \to \cW$, where $\cal Z$ is the receiver observation space until the decoding time $T$, i.e., $Y^{T}$ takes values in $\cZ$. The decoding time $T$ should be a stopping time\footnote{A discussion of stopping times can be  found in~\cite[sect.~10.8]{Wil91}.}  with respect to the receiver observation $Y^n$ otherwise the decision of when to decode would depend on future channel outputs and the decoder would no longer be causal.   We treat the case when $\EX{T} < \infty$, and point out that what we are  setting out to prove, namely (\ref{eq:whatweprove}), is trivially true when $\EX{T} = \infty$.

The codes we consider here differ from non-block (also called sequential) codes with variable delay, such as those studied in \cite{Hor63} and~\cite{Kud79}. In sequential coding, the message (typically as an infinite stream of bits) is introduced to the transmitter and decoded by the receiver in a progressive fashion.  Delay is measured separately for each bit, and is defined as the time between the introduction and decoding of the bit. This is in contrast to the codes considered in this paper, where the entire message is introduced to the transmitter at the start of communication, and $T$ measures the duration of the communication.  Due to their different problem formulation, sequential codes with feedback have reliability functions that differ from those for variable-length block codes, just as fixed constraint length convolutional codes have reliability functions that differ from those of fixed-length block codes.

The observation space $\cal Z$ is a collection of channel output sequences and for a DMC with feedback the length of these sequences may vary. (The length of the channel input itself may depend on the  channel realization). Nevertheless, these sequences have the property of being prefix-free (otherwise the decision to stop would require knowledge of the future).  Thus, $\cZ$ can be represented as the leaves of a complete $|\cY|$-ary tree $\cT$ (complete in the sense that each intermediate node has $|\cY|$ descendants), and has expected depth $\EX{T} < \infty$. Note that the decision time $T$ is simply the first time the sequence $Y_1,Y_2,\dots$ of channel outputs hits a leaf of $\cT$. Furthermore we may label each leaf of $\cal T$ with the message decoded  by the receiver when that leaf is reached. This way the decoder is completely specified by the labeled tree $\cal T$. The message statistics, the code, and the transition probabilities of the channel determine a probability measure on the tree $\cal T$.

\section{Binary Hypothesis Testing with Feedback} 
\label{sec:prelim} 
The binary case ($M=2$) will play a key role in our main proof. In this section we assume that the message set contains only two elements.  We will arbitrarily denote the two hypotheses by $A$ and $N$ (ACK and NACK, respectively). We denote by $\PRM{A}$ and $\PRM{N}$ the corresponding probability distributions on the leaves of $\cal T$.

The following proposition bounds the Kullback-Leibler divergence $D(\PRM{A}\|\PRM{N})$. It will be used in the main result of this section to bound the error probability of binary hypothesis testing with feedback. The reader familiar with Stein's Lemma will not be surprised by the fact that the Kullback-Leibler divergence $D(\PRM{A}\|\PRM{N})$ plays a key role in binary hypothesis testing with feedback. The steps here closely parallel those in \cite[Sec. III] {TchTel05}  and \cite[Sec. 2.2]{CsiShi04}.

\begin{proposition} \label{prop:expkl} 
For any binary hypothesis testing scheme
  for a channel with feedback \[D(\PRM{A}\|\PRM{N})\leq \DX \ECX{T}{A}  \]
  where $T$ is the decision stopping time, $\EX{T} < \infty$, and
  $\ECX{T}{A}$ denotes the expectation  of $T$  conditioned on hypothesis $A$.
%I deleted the explanation using $\PRM{A}$ as probability measures. It was correct
%but confusing. 
\end{proposition} 

\begin{proof} In the
	following, we will denote probability under hypothesis $A$ by $\PrA(\cdot)$
	and probability under hypothesis $N$ by $\PrN(\cdot)$.  Let
	\begin{equation} V_n=\ln\dfrac{\PrA(Y_1,\dots,Y_n)}{\PrN(Y_1,\dots,Y_n)},
	\end{equation} so that $ D(\PRM{A}\|\PRM{N})=\ECX{V_{T}}{A}$, and the
	proposition is equivalent to the statement $\ECX{V_{T} -\DX T}{A}\leq 0$.
	Observe that 
\begin{equation} V_n=\sum_{k=1}^n U_k \quad\text{where
	  $U_k=\ln\dfrac{\PrA(Y_k|Y^{k-1})}{\PrN(Y_k|Y^{k-1})}$.} 
\end{equation}
Note now that 
\begin{align} \ECX{U_k}{A,Y^{k-1}} &=\ECX{\ln \frac{
		\PrA(Y_k|Y^{k-1})} {\PrN(Y_k|Y^{k-1})}}{A,Y^{k-1}}  \nonumber\\
		&=\ECX{\ln \frac{
		\PrA(Y_k|X_k=f_k(A,Y^{k-1}),Y^{k-1})}{\PrN(Y_k|X_k=f_k(N,Y^{k-1}),Y^{k-1})}}{A,
		X_k=f_k(A,Y^{k-1}), Y^{k-1}} \nonumber\\ &=\sum_{y \in \cY}
		\PCX{Y_k=y}{X_k=f_k(A,Y^{k-1})}  \ln\frac{
		\PCX{Y_k=y}{X_k=f_k(A,Y^{k-1})} }{ \PCX{Y_k=y}{X_k=f_k(N,Y^{k-1})} }
		\nonumber\\  &\leq \DX, 
\end{align} 
where $f_k(\cdot,\cdot)$ is the
		encoder function at time $k$.  Consequently,  $\{V_n-n\DX\}$ is a
		supermartingale under hypothesis $A$. Observe that the existence of a
		$\lambda > 0$ for which $p(y|x) > \lambda$ for all $x,y$ implies that
		$|U_k| < \ln \frac1\lambda$. We can now use Doob's Optional-Stopping
		Theorem (see e.g.~\cite[Sec.~10.10]{Wil91}) to conclude that $\ECX{V_T-\DX T}{A}
		\leq 0$.  \end{proof}

We can apply Proposition~\ref{prop:expkl} to find a lower bound on the error
probability of a binary hypothesis testing problem with feedback. The bound is
expressed in terms of the expected decision time. 

\begin{lemma} \label{prop:pe-bound} The error probability of a binary
  hypothesis test performed across a DMC with feedback and variable-length
  codes  is lower bounded by \[ \Pe \geq \frac{\min\{p_A,p_N\}}{4}
  e^{-\CX_1\EX{T}}  \] where $p_A$ and $p_N$ are the \latin{a priori}
  probabilities of the hypotheses.  
\end{lemma}

\begin{proof} Each decision rule corresponds to a tree where each
  leaf $Y^{T}$  is associated with a decoded hypothesis $\hat{W}(Y^{T})$. Thus
  we can partition the leaves into two sets corresponding to the two
  hypotheses.  \begin{align*} \cS &= \{ y^T : \hat{W}(y^T) = A \} \\ \bar{\cS}
	&= \{ y^T : \hat{W}(y^T) \ne A \} \end{align*} where $\cS$ is the decision
	region for hypothesis $A$.

The log sum inequality \cite{CovTho91,CsK81} (or data processing lemma for
divergence) implies \begin{equation} \label{eq:kl-logsum}
  \KLD{\PRM{A}}{\PRM{N}} \geq \PRM{A}(\cS) \ln
  \frac{\PRM{A}(\cS)}{\PRM{N}(\cS)} + \PRM{A}(\bar{\cS}) \ln
  \frac{\PRM{A}(\bar{\cS})}{\PRM{N}(\bar{\cS})}. \end{equation} By
  Proposition~\ref{prop:expkl}, $\DX \ECX{T}{A} \ge \KLD{\PRM{A}}{\PRM{N}}$,
  thus~(\ref{eq:kl-logsum}) can be re-arranged to give \begin{equation}
	\label{eq:depth_conda} \DX \ECX{T}{A}\geq - \PRM{A}(\cS) \ln \PRM{N}(\cS)
	-\binh(\PRM{A}(\bar{\cS})), \end{equation} where $\binh(\cdot)$ is the
	binary entropy function. Writing the overall probability of error in terms
	of marginal error probabilities yields \[\Pe =p_N \PRM{N}(\cS)+ p_A
	\PRM{A}(\bar{\cS}), \] which allows us to bound $\PRM{N}(\cS)$ as \[
	\PRM{N}(\cS)\leq \frac{\Pe}{p_N} \leq \frac{\Pe}{\min\{p_A,p_N\}}. \] Substituting back
	into~(\ref{eq:depth_conda}) yields a bound on the expected depth of the
	decision tree conditioned on $A$ just in terms of $\PRM{A}$ and the
	\latin{a priori} message probabilities \begin{equation}
	  \label{eq:treedeptha} \DX \ECX{T}{A} \geq -\PRM{A}(\cS) \ln \frac{\Pe}
	  {\min \{p_A,p_N\}}- \binh(\PRM{A}(\bar{\cS})).  \end{equation}
Following identical  steps with the roles of $A$ and $N$ swapped yields
\begin{equation} \label{eq:treedepthr} \DX \ECX{T}{N} \geq -\PRM{N}(\bar\cS)
  \ln \frac{\Pe} {\min \{p_A,p_N\}}- \binh(\PRM{N}(\cS)).  \end{equation}
 We  can now average both sides of~(\ref{eq:treedeptha}) and~(\ref{eq:treedepthr})
  by weighting with the corresponding \latin{a priori} probabilities. If we do
  so and use the facts that $p_A\PRM{A}(\cS)+p_N\PRM{N}(\bar\cS)$ is the
  probability of making the correct decision and $p_A\PRM{A}(\bar\cS)+p_N\PRM{N}(\cS)$ 
is the probability of making an error together with the concavity of the 
binary entropy function, we obtain the following
  unconditioned bound on the depth of the decision tree \begin{align*} \DX
	\EX{T} &\geq -(1-\Pe) \ln \frac{\Pe}{\min\{p_A,p_N\}}-  \binh(\Pe) \\ &\geq
	-\ln \Pe-2\binh(\Pe)+\ln\min\{p_A,p_N\}\\ &\ge -\ln \Pe -2\ln 2 +
	\ln\min\{p_A, p_N\}.  \end{align*}

Solving for $\Pe$ completes the proof.
\end{proof} It is perhaps worthwhile pointing out why the factor
$\min \{p_A,p_N \}$ arises:  if one of the hypotheses has small \latin{a
priori} probability, one can achieve an equally small error probability by
always deciding for the other hypothesis, irrespective of the channel
observations. 

\section{Expected Tree Depth and Channel Capacity} 
\label{sec:chan_ca}
Given the channel observations $y^n$, one can calculate the \latin{a posteriori}
probability $p_{W|Y^n}(w|y^n)$ of any message $w \in \cW$.  Recall that a
maximum \latin{a posteriori} (MAP) decoder asked to decide at time $n$ when
$Y^n=y^n$ will chose (one of) the message(s) that has the largest \latin{a
posteriori} probability $p_{\max}=\max_w p_{W|Y^n}(w|y^n)$.  The probability of
error will then be $\Pe(y^{n})=1-p_{\max}$. Similarly, we can define the
probability of error of a MAP decoder for each leaf of the observation tree
$\cal T$. Let us denote by $\Pe(y^{T})$ the probability of error given the
observation $y^{T}$. The unconditioned probability of error is then
$\Pe=\EX{\Pe(Y^{T})}$. 

For any fixed $\delta>0$ we can define a stopping time $\tau$ as the first time that the error
probability goes below $\delta$, if this happens before $T$, and as $T$
otherwise: \begin{equation}\label{eq:tau_defn} \tau = \inf \left\{n:
  \left(\Pe(y^n)\leq \delta\right) \text{ or } ( n = T ) \right\} \end{equation}

If $\Pe(Y^{\tau})$ exceeds $\delta$, then we are certain that  $\tau = T$, and $\Pe(Y^n) > \delta$ for all $0\le n \le T$, so the event $\Pe(Y^\tau) > \delta$ is included in the event $\Pe(Y^T) > \delta$. (We have inclusion instead of equality since   $\Pe(Y^\tau) \leq \delta$ does not exclude $\Pe(Y^T) > \delta$.)   Thus
\begin{equation}
  \label{eq:peexdelta} \PX{\Pe(Y^{\tau}) > \delta} \leq \PX{\Pe(Y^{T})>\delta}
  \leq \frac{\Pe}{\delta}, 
\end{equation}
where the second inequality is an application of Markov's inequality.

Given a particular realization $y^n$ we will denote the entropy of the 
\latin{a posteriori} distribution $p_{W|Y^n}(\cdot|y^n)$ as $\h{y^n}$. Then $\HX{n}$ is a random variable\footnote{Notice that $\h{y^n}$ is commonly written as  $H(W|Y^{n}=y^{n})$. We cannot use the standard notation since it becomes problematic when we substitute $Y^n$ for $y^n$ as we just did.} and 
$\EX{\HX{n}} = H(W|Y^n)$. If $\Pe(y^{\tau})\leq \delta \leq \frac12$, then from Fano's inequality it follows that  
\begin{equation}
  \label{eq:htau_bound} \h{y^\tau} \leq \binh(\delta) +\delta \ln M.
\end{equation}
The expected value of  $\HX{\tau}$ can be bounded by conditioning on the event
$\Pe(Y^{\tau})\leq \delta$ and its complement then
applying~(\ref{eq:htau_bound}) and then~(\ref{eq:peexdelta}) as follows
\begin{align*} \EX{\HX{\tau}} &= \ECX{\HX{\tau}}{\Pe(Y^{\tau})\leq
  \delta}\PX{\Pe(Y^{\tau})\leq \delta}+  \ECX{\HX{\tau}}{\Pe(Y^{\tau})>
  \delta}\PX{\Pe(Y^{\tau})> \delta} \\ &\leq \left(\binh(\delta) +\delta \ln M
  \right)\PX{\Pe(Y^{\tau})\leq \delta} + (\ln M) \PX{\Pe(Y^{\tau})> \delta)} \\
  &\leq \binh(\delta)  + \left( \delta+\frac{\Pe}{\delta} \right) \ln M.
\end{align*}
This upper bound on the expected posterior entropy at time $\tau$ can be turned
into a lower bound on the expected value of $\tau$ by using the channel
capacity as an upper bound to the expected change of entropy. This notion is
made precise by the following lemma, \begin{lemma} \label{prop:exp_tau} For any
  $0 <\delta \leq \frac12$ \[ \EX{\tau} \ge
  \left(1-\delta-\frac{\Pe}{\delta}\right)\frac{ \ln
  M}{\CX}-\frac{\binh(\delta)}{\CX}. \] \end{lemma} \begin{proof} Observe that
	$\{\HX{n}+n \CX\}$ is a submartingale (an observation already made
	in~\cite[Lemma 2]{Bur76}). To see this, 
\begin{align*}
	  \ECX{\HX{n}-\HX{n+1}}{Y^{n}=y^{n}} 
&=I(W;Y_{n+1}|Y^{n}=y^{n})\\
&\stackrel{(a)}{\leq}I(X_{n+1};Y_{n+1}|Y^{n}=y^{n})\\ 
&\leq \CX 
\end{align*}
	  where $(a)$ follows from the data processing inequality and the fact that
	  $W$---$X_{n+1}$---$Y_{n+1}$ forms a Markov chain given  $Y^{n}=y^{n}$.
Hence  $\{\HX{n}+n \CX\}$ is indeed a submartingale. Since $\h{y^n}$ is bounded
between $0$ and $\ln M$ for all $n$, and the expected stopping time $\EX{\tau}
\le \EX{T} < \infty$, Doob's Optional-Stopping Theorem allows us
to conclude that at time $\tau$ the expected value of the submartingale must be
greater than or equal to the initial value, $\ln M$. Hence 
\begin{align*} 
\ln M = \HX{0}  &\leq   \EX{\HX{\tau}+\tau \CX} \\ &= \EX{\HX{\tau}}+\EX{\tau} \CX\\ &\le
  \binh(\delta) + \left( \delta + \frac\Pe\delta \right) \ln M + \EX{\tau} \CX.
\end{align*} Solving for $E[\tau]$ yields \[ \EX{\tau} \ge
\left(1-\delta-\frac{\Pe}{\delta}\right) \frac{ \ln
M}{\CX}-\frac{\binh(\delta)}{\CX}. \] \end{proof}

\section{Burnashev's Lower Bound} 
\label{sec:burnbound}

In this section we will combine the two bounds we have established in the
preceding sections to obtain a bound on the overall expected decoding time.
Lemma~\ref{prop:exp_tau} provides a lower bound on $\EX{\tau}$ as a function of
$M$, $\delta$ and $\Pe$. We will show that a properly constructed binary
hypothesis testing problem allows us to  use Lemma~\ref{prop:pe-bound} to lower bound
the probability of error in terms of $\ECX{T-\tau}{Y^{\tau}}$. This in turn
will lead us to the final bound on $\EX{T}$.

The next proposition states that a new channel output symbol can not change the
\latin{a posteriori} probability of any particular message by more than some constant factor when $\DX$
is finite.
\begin{proposition} \label{prop:postent} $C_1 < \infty$ implies 
\[\lambda p\left(w|y^{n-1}\right)    \leq  p\left(w|y^n\right)  \leq \frac{ p\left(w|y^{n-1}\right)}{\lambda}, \] 
where  $0< \lambda=\min_{x,y} p(y|x) \le \frac12$.  \end{proposition}

\begin{proof} Using Bayes' rule, the posterior may be written recursively as
\[ 
p\left(w|y^n\right) = p\left(w|y^{n-1}\right) \frac{p\left(y_n|x_n=f_n(w,y^{n-1})\right)}{ p(y_n|y^{n-1})}.
\]
The quotient may be upper and lower bounded using $1 \ge p\left(y_n|x_n=f_n(w,y^{n-1})\right) \ge \lambda $ and $1 \ge p(y_n|y^{n-1}) \ge \lambda$, which yields the statement of the proposition.  \end{proof}

Our objective is to lower bound the probability of error of a decoder that decides at time $T$. The key idea is that a binary hypothesis decision such as deciding whether or not $W$ lies in  some set $\cG$ can be made at least as reliably as a decision on the value of  $W$ itself.

Given a set $\cG$ of messages, consider deciding between $W\in \cG$ and $W\not\in \cG$ in the following way: given access to the original decoder's estimate $\hat W$, declare that $W
\in \cG$ if $\hat W\in\cG$, and declare $W \not\in\cG$ otherwise. This binary
decision is always correct when the original decoder's estimate $\hat W$ is
correct.  Hence the probability of error of this (not necessarily optimal)
binary decision rule cannot exceed the probability of error of the original decoder, for
any set $\cG$. Thus the error probability of the optimal decoder deciding at time 
$T$ whether or not $W\in\cG$ is a lower bound to the error probability of any 
decoder that decodes $W$ itself at time $T$. This fact is true even if the set $\cG$ is chosen at a 
particular stopping time $\tau$ and the error probabilities we are calculating 
are  conditioned on the observation $Y^{\tau}$.

For every realization of $Y^{\tau}$, the message set can be divided into two
parts, $\cGt$ and its complement $\cW\setminus\cGt$,  in such a way that both
parts have an \latin{a posteriori} probability greater than $\lambda\delta$.
The rest of this paragraph describes how this is possible. 
%The proof, which is not particularly enlightening, may be initially skipped.
From the definition of $\tau$, at time $\tau-1$ the \latin{a posteriori} probability of every
message is smaller than $1-\delta$. This implies
that the sum of the \latin{a posteriori} probabilities of any set of $M-1$
messages is greater than $\delta$ at time $\tau -1$, and by
Proposition~\ref{prop:postent}, greater than $\lambda\delta$  at time $\tau$.
In particular, $\Pe(y^{\tau}) \geq \lambda \delta$. We separately consider the
cases $\Pe(y^{\tau}) \leq  \delta$ and $\Pe(y^{\tau}) > \delta$. In the first
case, $\Pe(y^\tau)\leq \delta$, let $\cGt$ be the set consisting of only
the message with the highest \latin{a posteriori} probability at time $\tau$. 
The \latin{a posteriori} probability of $\cGt$ then satisfies 
$\PX{\cGt} \geq 1-\delta \geq 1/2 \geq \lambda\delta$. As
argued above, its complement (the remaining $M-1$ messages) also has \latin{a posteriori}
probability greater than $\lambda\delta$, thus for this $\cGt$,
$\PCX{\cGt}{Y^{\tau}} \in [\lambda\delta,1-\lambda\delta]$. 
In the second case, namely when $\Pe(y^\tau)>\delta$, the
\latin {a posteriori} probability of each message is smaller than $1-\delta$.
In this case the set $\cGt$ may be formed by starting with the empty set and
adding messages in arbitrary order until the threshold $\delta/2$ is exceeded. This
ensures that the \latin{a priori} probability of $\cGt$ is greater than 
 $\lambda\delta$. Notice that the threshold will be
exceeded by at most $1-\delta$, thus the complement set has an \latin{a
posteriori} probability of at least $\delta/2 > \lambda\delta$. Thus 
$\PCX{\cGt}{Y^{\tau}} \in [\lambda\delta,1-\lambda\delta]$.

%Now we apply Lemma~\ref{prop:pe-bound} to the binary hypothesis testing problem that starts at time $\tau$ and decides at time $T$

For any realization of $Y^{\tau}$ we have the binary hypothesis testing problem, running from $\tau$ until $T$, 
deciding  whether or not $W\in\cGt$. Notice that the \latin{a priori} probabilities
 of the two hypotheses of this
binary hypothesis testing problem are the \latin{a posteriori} probabilities of
$\cGt$ and $\cW\setminus\cGt$ at time $\tau$ each of which is shown to be greater
than $\lambda \delta$ in the paragraph above . We apply Lemma~\ref{prop:pe-bound}
with $A=\cGt$ and $N=\cW\setminus\cGt$ to lower bound the probability of error of
the binary decision made at time $T$ and, as argued above, we use the result to
lower bound the probability that $\hat W\ne W$. Initially everything is
conditioned on the channel output up to time $\tau$, thus 
\[\PCX{\hat{W}\left(Y^{T}\right)\neq W }{Y^{\tau}} \geq \frac{\lambda\delta}{4} e^{-\DX
\ECX{T-\tau}{Y^{\tau}}}. \]
Taking the expectation of the above expression over all realizations of $Y^\tau$ 
yields the unconditional
probability of error \[\Pe =\EX{\PCX{\hat{W}\left(Y^{T}\right)\neq W }{Y^{\tau}}}\geq \EX{
\frac{\lambda\delta}{4} e^{-\DX \ECX{T-\tau}{Y^{\tau}}}   } .\] Using the
convexity of $e^{-x}$ and Jensen's inequality, we obtain
\[\Pe  \geq \frac{\lambda\delta}{4} e^{-\DX \EX{T -\tau}}. \]
Solving for $\EX{T-\tau}$ yields \begin{equation} \label{eq:boundon2} \EX{T
  -\tau} \geq \frac{-\ln \Pe - \ln 4 +\ln (\lambda \delta) }{\DX}.
\end{equation} Combing Lemma~\ref{prop:exp_tau} and~(\ref{eq:boundon2}) yields:
\begin{theorem} 
  The expected decoding time $T$ of any variable-length block code for
  a DMC used with feedback is  lower bounded by  \begin{equation} \EX{T} \geq
	\left(1-\delta-\frac{\Pe}{\delta}\right)\frac{ \ln M}{\CX} + \frac{-\ln
	\Pe}{\DX}  -\frac{\binh(\delta)}{\CX} + \frac{\ln (\lambda \delta) - \ln 4
	}{\DX}, \end{equation} where $M$ is the cardinality of the message set,
	$P_e$ the error probability, $\lambda = \min_{x\in\cX,y\in\cY} p(y|x)$, and
	$\delta$ is any number satisfying $0<\delta \leq \frac{1}{2}$.  \hfill\QED
  \end{theorem} 
  
  Choosing the parameter $\delta$ as $\delta=-\frac{1}{\ln P_e}$ achieves the required scaling
  for (\ref{eq:whatweprove}).

\section{Summary}\label{sec:conc} We have presented a new derivation of
Burnashev's asymptotically tight lower bound to the average delay needed for a
target error probability when a message is communicated across a DMC used with
(channel output) feedback.  Our proof is simpler than the original, yet
provides insight by clarifying the role played by the quantities that appear in
the bound.
Specifically, from the channel coding theorem we expect it to take roughly
$\frac{\ln M}{C}$ channel uses to reduce the probability of error of a MAP
decision to some small (but not too small) value.  At this point we can
partition the message set in two subsets, such that  neither subset has too
small an \latin{a posteriori} probability. From now on it takes
(asymptotically)  $-\frac{\ln P_e}{C_1}$ channel uses to decide with
probability of error $P_e$ which of the two sets contains the true message. It
takes at least as many channel uses to decide which message was selected and
incur  the same error probability.

For obvious reasons we may call the two phases the communication and the binary
hypothesis testing phase, respectively. These two phases  exhibit a pleasing
similarity to the communication and confirmation phase of the optimal scheme
proposed and analyzed by Yamamoto and Itoh in \cite{YamIto79}. The fact that
these two phases play a key role in proving achievability as well as in proving
that one cannot do better suggests that they are an intrinsic component of an
optimal communication scheme using variable-length block codes over DMCs with
feedback.

\section*{Acknowledgement} The authors would like to thank R.~G.~Gallager for
his help in pointing out an error in an earlier version of this paper and the
reviewers for their helpful comments.

\bibliographystyle{ieeetr}

\end{document}